\documentclass[preprint,showpacs,preprintnumbers,amsmath,amssymb]{revtex4}
\usepackage{graphicx}
\usepackage{floatflt}
\usepackage{dcolumn}
\usepackage{bm}
\usepackage{rotating}
\usepackage{natbib}
\usepackage{txfonts}

\begin{document}
   \title{Equations of state and stability of 
color-superconducting \\ quark matter cores in hybrid stars }

   \author{B. K. Agrawal}
              \email{bijay.agrawal@saha.ac.in}
   \affiliation{Saha Institute of Nuclear Physics, Kolkata - 700064, India.}

\date{\today}

  \begin{abstract}

The stable configurations of non-rotating and rotating hybrid stars
composed of colour superconducting quark matter core are constructed
using several  equations of state (EOSs).  We use a set of diverse
EOSs for the nuclear matter which represents the  low density phase.
The EOSs at higher densities correspond to  the quark matter in the
colour superconducting phase and are computed within the NJL-like model
for different values of the scalar diquark and vector current couplings
strengths. The phase transition to the quark matter is computed by
a Maxwell construction.  We find that the stability of the hybrid
stars are mainly governed by the behaviour of the EOSs for the colour
superconducting quark matter. However the compositions of hybrid star are
sensitive to the EOS of the nuclear matter. The value of the critical
rotation frequency for the hybrid star depends strongly on the EOS of
the nuclear matter as well as that for the colour superconducting quark
matter.  Our results indicate that the EOS for the colour superconducting
quark matter can be obtained, by adjusting the parameters of the NJL
model, to yield the stable configurations of the hybrid star having the
maximum mass $\sim 1.5M_\odot$ in the non-rotating limit and the critical
rotation frequency $\sim 1$ kHz.

\end{abstract}

\maketitle

\section{Introduction}
\label{intro-sec}

The present knowledge of quantum chromodynamics (QCD) suggests
that quark matter might be in different  color superconducting phases
at high densities. Thus, one expects the core of the  hybrid stars to be
composed of color superconducting quark matter (CSQM) surrounded
by a nuclear mantle.  The possible CSQM phases are the two-flavor color
superconductor (2SC) \cite{Alford98a,Ruester04,Ruester05}, the colour
flavour locked (CFL) phase \cite{Alford99a,Rajagopal01}, and crystalline
color superconductor (CCS) \cite{Alford01a,Rajagopal06}. The speculation
that the CSQM exists in the core of the hybrid stars has triggered many
theoretical investigations both on the modeling of the equation of state
(EOS) of quark matter and on the phenomenological signatures of the
presence of quark matter in the compact stars \cite{Alford08}.

The nuclear matter phase of the hybrid star is described by the
various models which can be broadly grouped into (i) non-relativistic
potential models \cite{Pandharipande75}, (ii) non-relativistic
mean-field models \cite{Chabanat97,Stone03,Mornas05,Agrawal06},
(iii) field theoretical based relativistic mean-field
models \cite{Prakash95,Glendenning99,Steiner05}
and (iv) Dirac-Brueckner-Hartree-Fock model
\cite{Muther87,Engvik94,Engvik96,Schulze06}.  The quark matter in the
colour superconducting phases are usually described either within the MIT bag
model or using a more realistic  NJL-like  model. The studies based on the
MIT bag model indicate the existence of  stable configurations of hybrid
stars with the CFL quark matter core
\cite{Alford03,Banik03,Drago04,Bombaci07}.  Further, the MIT bag model
predicts the absence of the 2SC colour superconducting phase in the
hybrid stars \cite{Alford02}.  The scenario is some what different when
NJL model is employed to study the hybrid stars with CSQM  core. 
The stable configurations of hybrid stars with 2SC quark matter core
are possible within the NJL model \cite{Blaschke05, Aguilera04, Aguilera05,
Grigorian04}.  However, earlier investigations
\cite{Baldo03,Buballa05,Klahn07} based on the NJL model ruled out
the possibility of CFL quark matter at the core of the hybrid stars,
because, it rendered the hybrid star unstable.  Only very recently
\cite{Pagliara08,Drago08}, it has been demonstrated that inclusion of
the six-fermion interaction term together with large enough values of
the scalar diquark coupling strength in the NJL model can yield stable
configurations of the hybrid star containing 2SC or CFL quark matter core.
The NJL model is also applied to study the possibility of existence of the
CCS quark matter phase in the hybrid stars \cite{Ippolito07,Ippolito08}.

The stability and the structure  of the non-rotating hybrid stars are
quite sensitive to the choice of the EOS of the nuclear matter and the
quark matter \cite{Alford03,Alford05}.  Further, one often finds that
even though the stable configurations of the non-rotating hybrid star
for a given EOS belong to the third family of compact stars, but,  the
maximum rotation frequency upto which these hybrid stars  are stable is
much lower than the corresponding mass-shedding (Keplerian) frequency
\cite{Banik05,Bhattacharyya05}.  The EOSs for the quark matter in the
unpaired or in the various colour superconducting phases  employed in
these investigations were obtained within the MIT bag model. Recently
\cite{Yang08}, a more realistic EOS for the unpaired quark matter computed
within the NJL model is used to show that the maximum mass of the
non-rotating hybrid stars depends sensitively on the choice of the EOS of
the nuclear matter. It is necessary to construct stable configurations
of the non-rotating and rotating hybrid stars using realistic EOSs for
the nuclear matter and for the  quark matter in the colour
superconducting phases.

In the present work, we compute several EOSs and use them to study
the properties of the non-rotating and rotating hybrid stars composed
of CSQM core.  The lower density part of these EOSs correspond to
the nuclear matter and are based on the variational and mean-field
approaches.  Our set of EOSs for the nuclear matter around the saturation
density ($\rho_0 = 0.16$ fm$^{-3}$) is constrained by the bulk properties
of the finite nuclei. But, their behaviour at densities,  $\rho > \rho_0$,
are significantly  different.  The EOSs for CSQM are calculated within
the NJL model using different values for the scalar diquark and vector
current coupling strengths. The EOS at intermediate densities are obtained
using a Maxwell construction.

The paper is organized as follows, in Sec. II we describe , in brief, the
models employed to construct the EOSs for nuclear matter and the CSQM.
In Sec. III we present  the results for the equilibrium sequences
for  non-rotating and rotating hybrid stars.  In Sec. IV we state our
conclusions.

{\section{Equations of state}}

\begin{figure}[ht]
 \centering
\resizebox{6.5in}{!}{\includegraphics[]{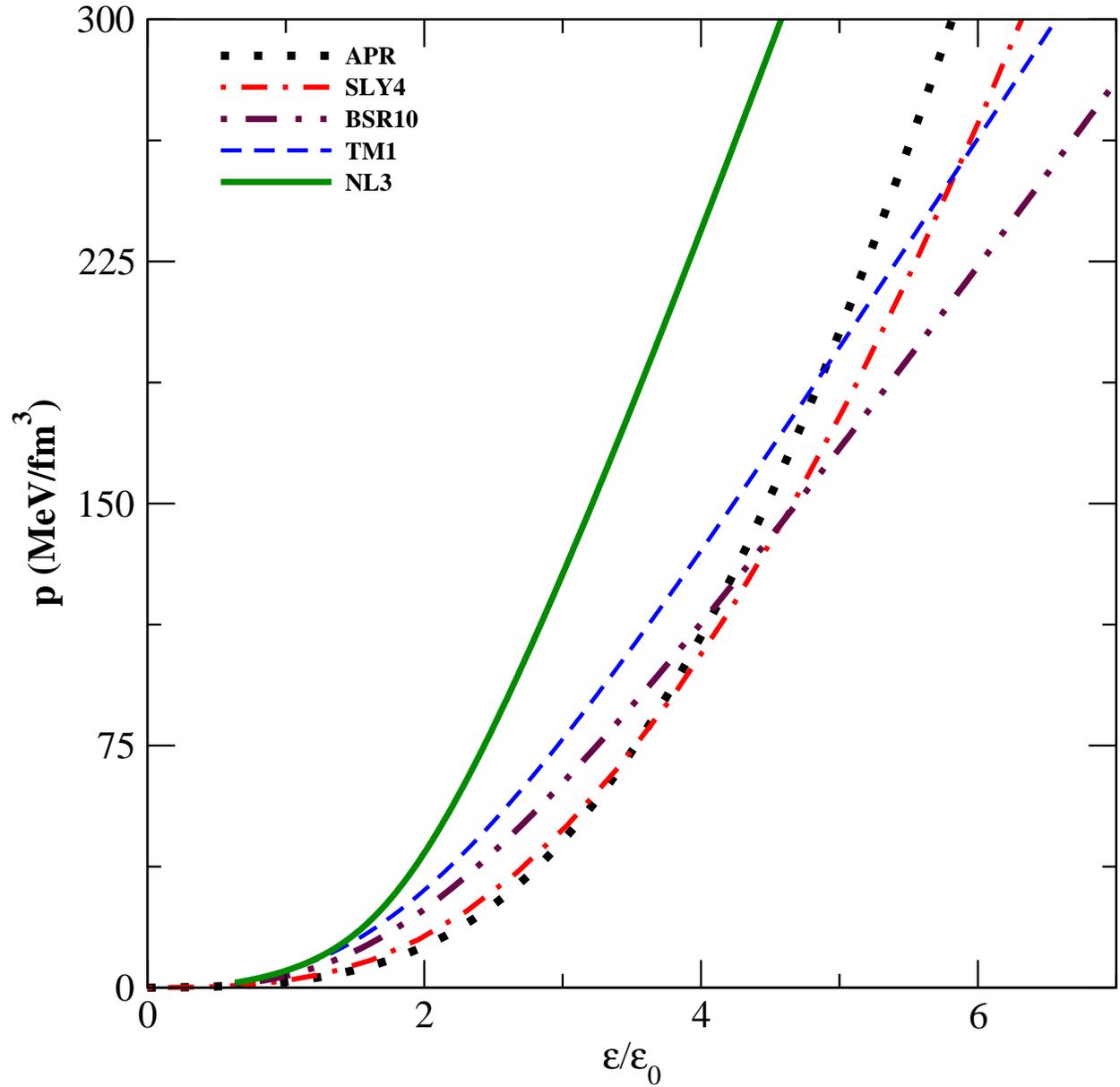}}
\caption{\label{fig:fig1} (Color online)
The nuclear matter EOSs plotted as pressure versus energy density. The
energy density is normalized by $\epsilon_0 = 150$ MeV/fm$^3$  which is
the typical value of the energy density for the nuclear matter at the
saturation density.  }

   \end{figure}

We compute the EOSs which correspond to the nuclear matter at lower
densities and CSQM in the 2SC or CFL phases at higher densities. The
EOS at intermediate densities are obtained using a Maxwell construction.
For nuclear matter in the $\beta$ equilibrium, we employ a set of
diverse EOSs  which are obtained using various approaches, like,
variational, non-relativistic mean field (NRMF) and relativistic mean
field (RMF).  In Fig. \ref{fig:fig1} we plot various nuclear matter EOSs.
The low density behaviour  these EOSs are very much similar as they
are constrained by the bulk properties of the finite nuclei. But, their
behaviour at higher densities are so different that the resulting neutron
star properties are at variance. In Table \ref{tab:tab1}, we list some
key properties of the non-rotating  neutron stars obtained using these
nuclear matter EOSs.  It can be seen from Table \ref{tab:tab1} that
the values of the maximum neutron star masses are in the range  of $2.0 -
2.8M_\odot$ and the radius $R_{1.4}$  at the canonical neutron star
mass vary between $11.3 - 14.8$km.  It is interesting to note that the
values of the maximum neutron star mass for both the APR and TM1 EOSs
are equal, but, the radius at the canonical mass of the neutron star
is reasonably smaller for the APR EOS. This is due to the fact that
the APR EOS is softer relative to the TM1 at intermediate densities
and it becomes stiffer at high densities as can be seen from
Fig. \ref{fig:fig1}. Similar is the case with SLY4 and BSR10 EOSs.
We shall see in next section  that these pairs of nuclear  matter
EOSs, for which the maximum neutron star masses are the same, yield
significantly different structure for the hybrid stars.

The EOSs for the CSQM in the 2SC or CFL phase
are obtained within the NJL model. 
The input variables of the NJL model are the chemical potentials for all
the quark flavours and colours in the chemical equilibrium which is  given
by the matrix:
 \begin{eqnarray}
\mu^{\alpha\beta}_{ab}= (\mu\delta^{\alpha\beta} +
\mu_QQ^{\alpha\beta}_f)\delta_{ab} + \left [\mu_3\left (T_3\right
)_{ab}+\mu_8\left (T_8\right )_{ab}\right ] \delta^{\alpha\beta},
\label{eq:chem_pot}
 \end{eqnarray} 
where, $\mu$ is the quark chemical potential, $\mu_Q$ is the chemical
potential of the electric charge equal to minus the electron chemical
potential $\mu_e$ and $\mu_3$ and $\mu_8 $ are the colour chemical
potentials associated with the two mutually commuting colour charges
of the $SU(3)_c$ gauge group. The explicit form of the electric charge
matrix  $Q_f = diag_f(\frac{2}{3}, -\frac{1}{3}, -\frac{1}{3})$, and for
the colour charge matrices $T_3 = diag_c(\frac{1}{2}, -\frac{1}{2}, 0)$,
and $\sqrt{3}T_8 = diag_c(\frac{1}{2}, \frac{1}{2}, -1)$.
In the mean-field approximation, the pressure at vanishing temperature
reads as,

\begin{equation}
p=4K\sigma_u\sigma_d\sigma_s -\frac{1}{4G_D}\sum_{c=1}^{3}\left
|\Delta_c\right |^2 -2G_S\sum_{\alpha=1}^{3}\sigma_\alpha^2
 +\frac{\omega_0^2}{4G_V}+ \frac{1}{{2\pi^2}} \sum_{i=1}^{18}\int_0^\Lambda
dk k^2 \left |\epsilon_i\right |+P_e  - B \label{eq:p_njl}
\end{equation} 
 where, $\sigma_{u,d,s}$  are the quark-antiquark
condensates and $\Delta_c$ are the three diquark condensates. 
 The values of $\sigma_i$ and $\Delta_c$ are determined using,
 \begin{eqnarray} 
\label{eq:sigma}
\frac{\partial p}{\partial \sigma_i}=0\\
\frac{\partial p}{\partial \Delta_c}=0. 
\label{eq:delta}
\end{eqnarray} 
In Eq. (\ref{eq:p_njl}) $\omega_0$ is the mean field expectation value for
isoscalar vector like meson $\omega$ given as \cite{Klahn07} $\omega_0 = 2G_V \langle
QM \mid \psi_u^\dagger \psi_u + \psi_d^\dagger \psi_d+\psi_s^\dagger
\psi_s \mid QM \rangle$.  
This field modifies also the chemical potentials: $\mu_{u,d,s} \rightarrow
\mu_{u,d,s} - \omega_0$. 
The $\epsilon_i$ are the dispersion relations
computed by following the Ref.  \cite{Ruester05}.  The
$\epsilon_i$ depend explicitly on the  values of current quark masses,
quark-antiquark and diquark condensates and various chemical potentials
appearing in Eq. (\ref{eq:chem_pot}).  The $P_e=\mu_e^4/(12\pi^2)$ is the
contribution to the pressure from the electrons.  The constant $B$ is so
determined that the pressure vanishes at zero density and temperature.
In addition to the Eqs. (\ref{eq:sigma}) and (\ref{eq:delta}), the
pressure must satisfy,
 \begin{eqnarray} n_Q \equiv \frac{\partial p}{\partial
\mu_Q} =0,\\ \label{eq:muQ}
  n_3 \equiv \frac{\partial p}{\partial \mu_3} =0,\\ \label{eq:mu3}
 n_8 \equiv \frac{\partial p}{\partial \mu_8} =0, \label{eq:mu8}
  \end{eqnarray}
so that local electric and colour charge neutrality conditions are met.
Once, the  pressure as a function of quark chemical potential is  known,
quark matter EOS can be easily computed.

The model parameters, the current quark masses $m_{u,d,s}$,
quark-antiquark coupling $G_S$, the strength $K$ of the six fermion or
"t Hooft" interaction and the cutoff parameter $\Lambda$ are taken to
be \cite{Rehberg96},

 \begin{eqnarray}
m_u = m_d = 5.5 \text{ MeV},\\
m_s=140.7 \text{ MeV},\\
G_S\Lambda^2=1.835,\\
K\Lambda^5=12.36,\\ 
\Lambda=602.3 \text{ MeV.}
\end{eqnarray}
After fixing the masses of the up and down quarks, $m_u =
m_d = 5.5 \text{ MeV}$, the other four parameters are chosen to reproduce
the following observables of vacuum QCD \cite{Rehberg96}: $m_\pi =
135.0\text{ MeV}$, $m_K = 497.7\text{ MeV}$, $m_{\eta'} = 957.8\text{
MeV}$, and $f_\pi = 92.4\text{ MeV}$. This parameter set gives
$m_\eta=514.8\text{ MeV}$.  The value of $B$ for this set of parameters
is $(425.4 \text{MeV})^4$.  There are two more parameters, the diquark
coupling strength  $G_D$ and the vector current coupling strength  $G_V$,
which are not known.  One expects that the diquark coupling has a similar
strength as the quark-antiquark coupling.  We construct quark matter
EOS for $G_D = 1.1 - 1.2 G_S$ with $G_V = 0 - 0.2 G_S $.

\begin{figure}[ht]
 \centering
\resizebox{6.5in}{!}{\includegraphics[]{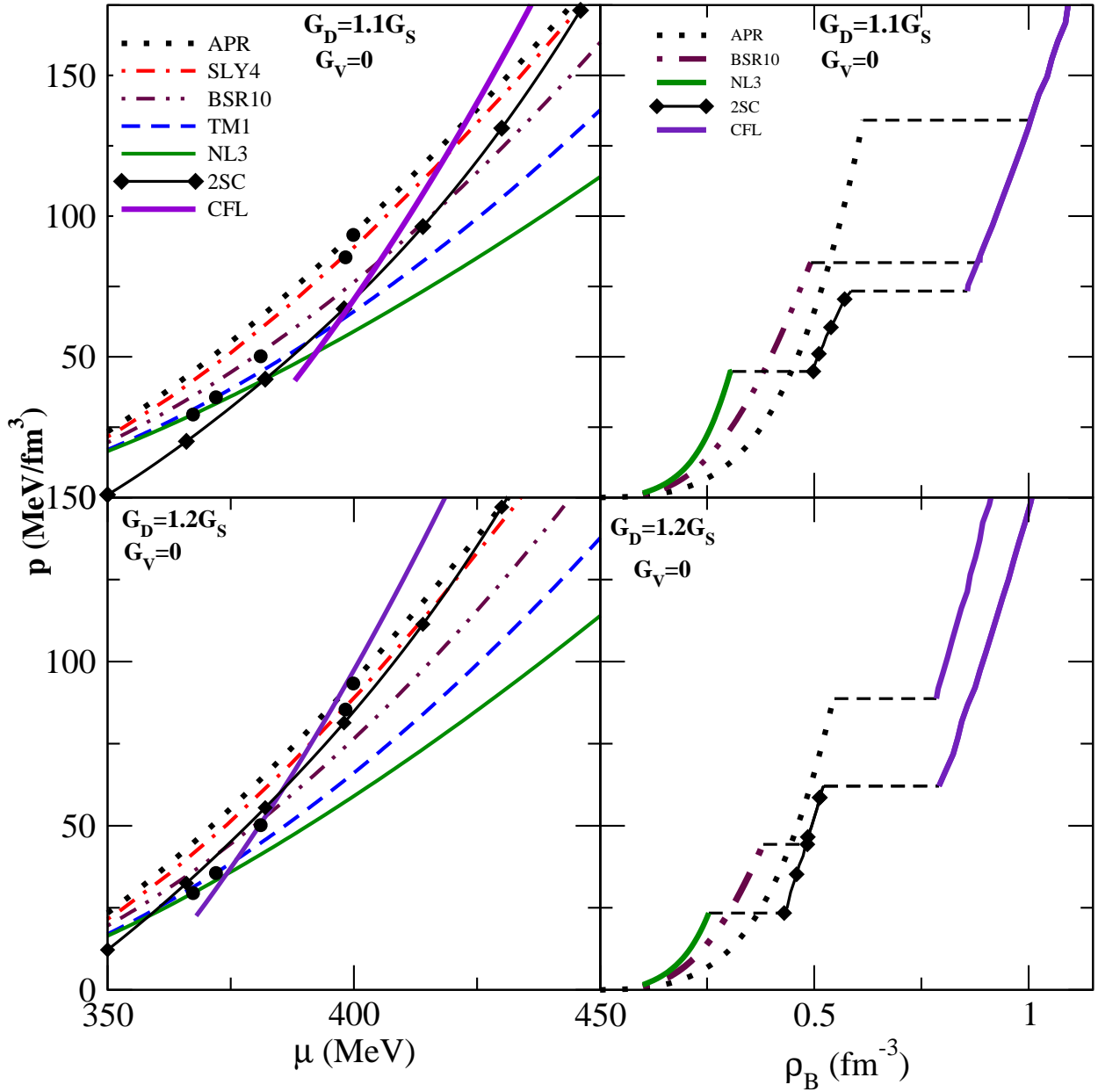}}
\caption{\label{fig:fig2} (Color online) The pressure as a function of
the quark chemical potential (left panel) and the baryon density (right
panel) for the nuclear and quark matter. The EOSs for the quark matter
in the 2SC and CFL phases are obtained within the NJL model using $G_V=0$
with $G_D=1.1G_S$ and $1.2G_S$. The solid circles on the various EOSs for
the nuclear matter indicate the pressure at the center of the neutron
star with the canonical mass ($1.4M_\odot$).  }
\end{figure}

\begin{figure}[ht]
 \centering
\resizebox{6.5in}{!}{\includegraphics[]{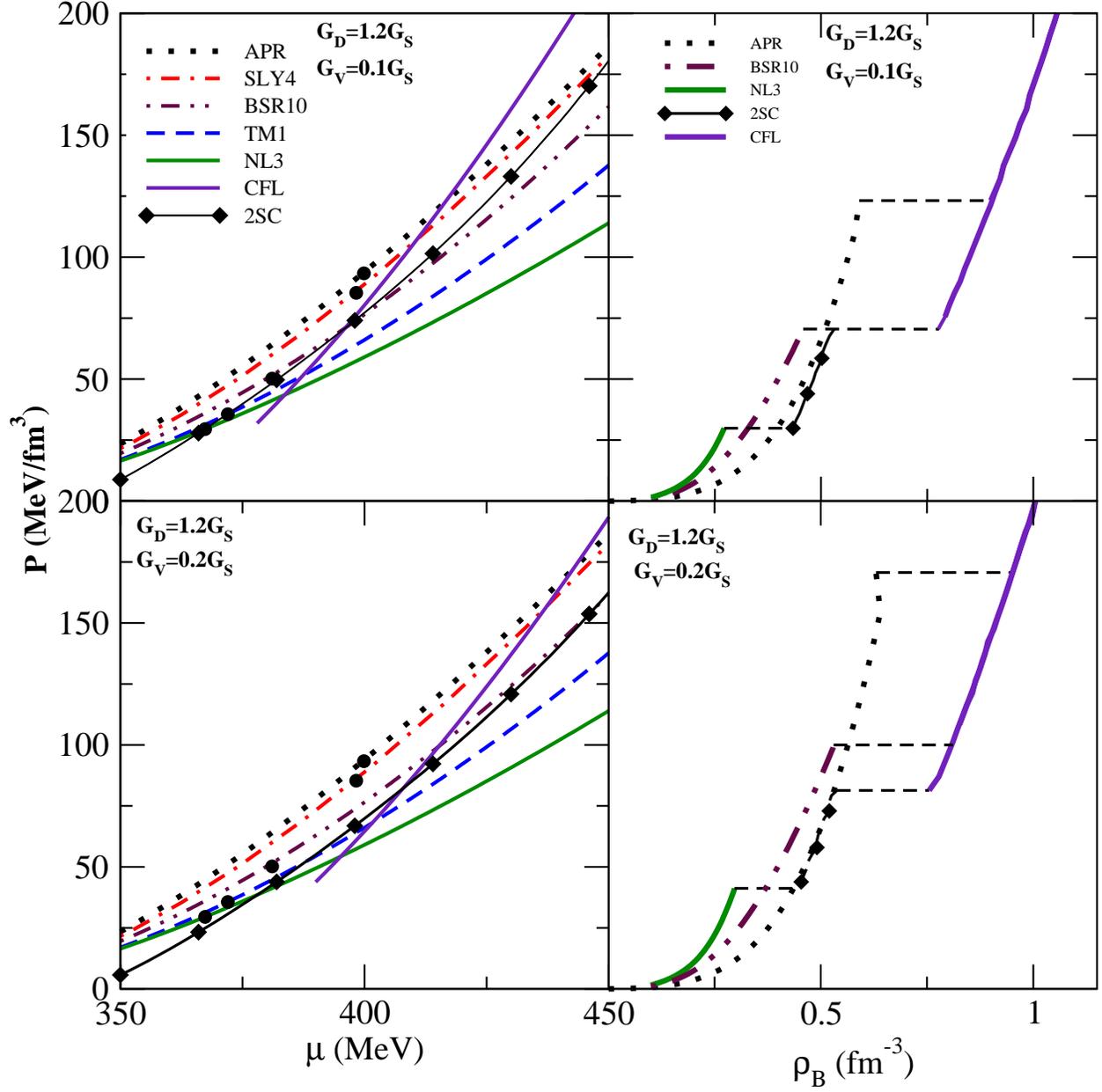}}
\caption{\label{fig:fig3} (Color online) Same as Fig. \ref{fig:fig2},
but, for $G_V = 0.1G_S$ and $0.2G_S$ with $G_D = 1.2G_S$.  }
   \end{figure}

In the 2SC phase, pairing occurs only between the $u$ and $d$ quarks and
the $s$ quarks remain unpaired leading to $\Delta_1 = \Delta_2 = 0$ and
$\Delta_3 \not= 0$. On the other hand, in the CFL phase,  $\Delta_1\not=
0$, $\Delta_2\not= 0$ and $\Delta_3\not = 0$.  In the left panels of
Figs. \ref{fig:fig2} and \ref{fig:fig3}, we plot the pressure as a
function of the quark chemical potential for the nuclear matter and for
the quark matter in the 2SC and CFL phases.  The phase realized at a
given chemical potential is the  one having largest pressure.  Thus, it
is evident from the $P - \mu$ curves that direct transition from nuclear
matter to the CFL quark matter occurs for the case of APR and SLY4 EOSs.
For the TM1 and NL3 EOSs, transition from nuclear matter to the CFL quark
matter proceeds via 2SC phase at intermediate densities.  For the BSR10
EOS, nuclear matter to CFL quark matter phase transition proceeds via
2SC phase only for $G_D=1.2G_S$ with $G_V=0$.  We see from these figures
that the pressure, at which the transition from the nuclear to quark
matter occurs,  decreases with increasing $G_D$ or decreasing $G_V$.
For instance, pressure at the phase transition reduces almost by a
factor of two with increase in $G_D$ from $1.1G_S$ to $1.2G_S$. The solid
circles on the various EOSs for the nuclear matter indicate the values of
$P_{1.4}$ which is the  pressure at the center of the neutron star with
the canonical mass ($1.4M_\odot$).  The values of transition pressure is
close to that of $P_{1.4}$ for the cases plotted in the lower and the
upper panels of Figs. \ref{fig:fig2} and \ref{fig:fig3}, respectively.
For the completeness, in the right panels of Figs. \ref{fig:fig2} and
\ref{fig:fig3}, we display the plots for the pressure as a function of
the baryon density for the case of APR, BSR10 and NL3 EOSs.  The phase
transition to the quark matter is computed by a Maxwell construction.

\section{Hybrid stars with CSQM core}

We construct the equilibrium sequences of the non-rotating and rotating
compact stars using the EOSs obtained in the last section.  These EOSs
correspond to the nuclear matter at lower densities and the CSQM in the
2SC or CFL phase at higher densities as shown in Figs. \ref{fig:fig2}
and \ref{fig:fig3}.  The nuclear matter EOSs are taken from the published
literature as summarized in Table \ref{tab:tab1}. The EOSs for the CSQM
are computed within the NJL model for different values of the scalar
diquark coupling strength $G_D$ and the vector current coupling strength
$G_V$. The other parameters of the model are determined by fit to some of
the observables of the vacuum QCD.  The various EOSs as obtained
in the present work can be completely specified by (i) the
source for the nuclear matter EOS as listed in Table \ref{tab:tab1} and
(ii) the values of $G_D$ and $G_V$ used in computing the EOS of the
CSQM within the NJL model.  The properties of spherically symmetric
non-rotating and axially symmetric rotating compact stars  are obtained by
solving the Einstein's equations in 1D and 2D, respectively. The numerical
computations are performed by using RNS code written by Stergioulas and
Friedman \cite{Stergioulas95}.

\begin{figure}[ht]
 \centering
\resizebox{6.5in}{!}{\includegraphics[]{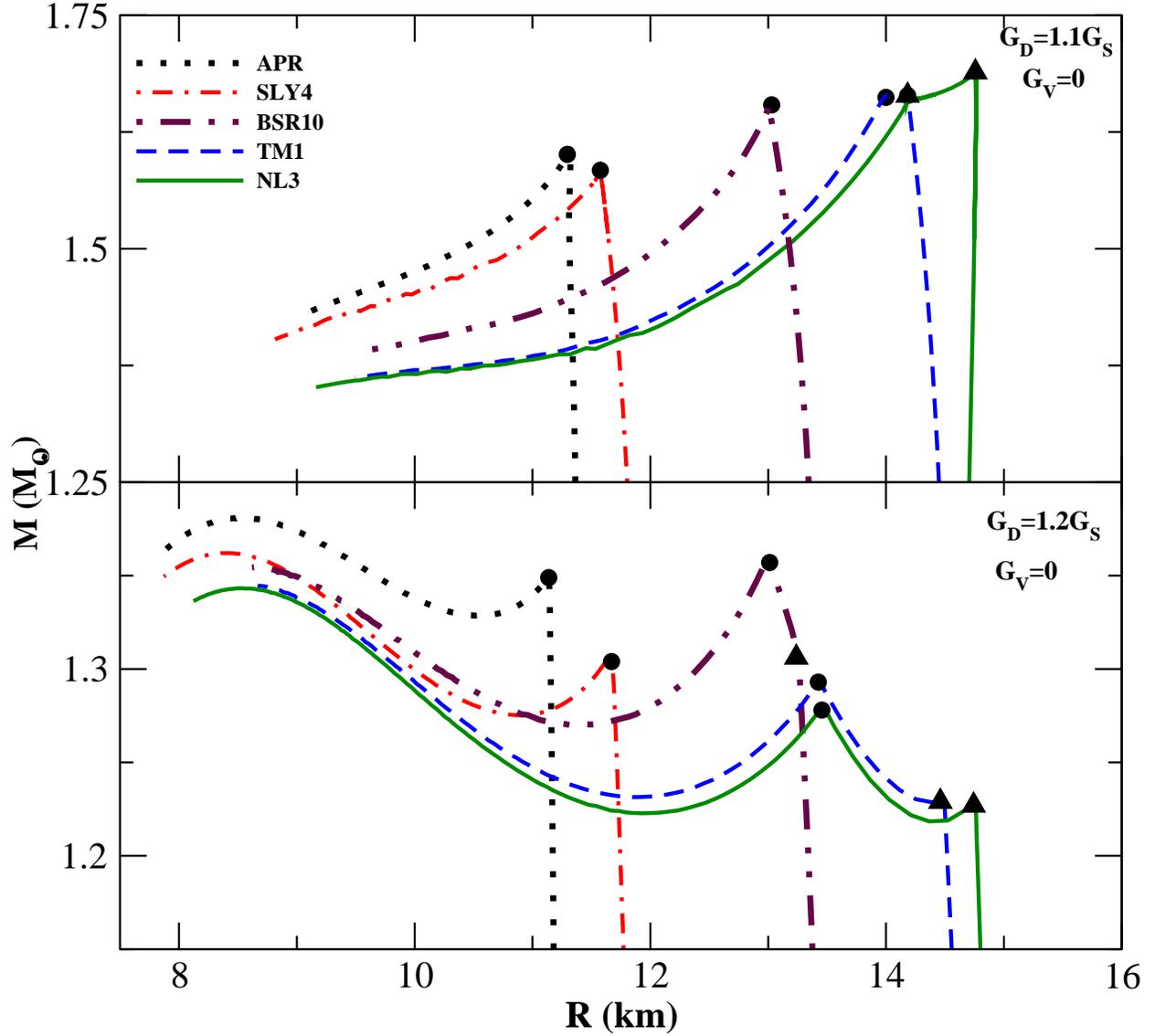}}
\caption{\label{fig:fig4} (Color online)
Plots for the mass-radius relationships for the  equilibrium sequences
of  non-rotating compact stars obtained using various EOSs as shown in
Fig. \ref{fig:fig2}. The curves on the left of the solid circles represent
the equilibrium sequences for the hybrid stars with core composed of
the quark matter in the CFL phase. The curves between the solid circles
and triangles represent the hybrid stars with 2SC quark matter core. The
absence of solid triangle on a curve means that the hybrid stars contain
quark matter only in the CFL phase.  }
   \end{figure}

\begin{figure}[ht]
 \centering
\resizebox{6.5in}{!}{\includegraphics[]{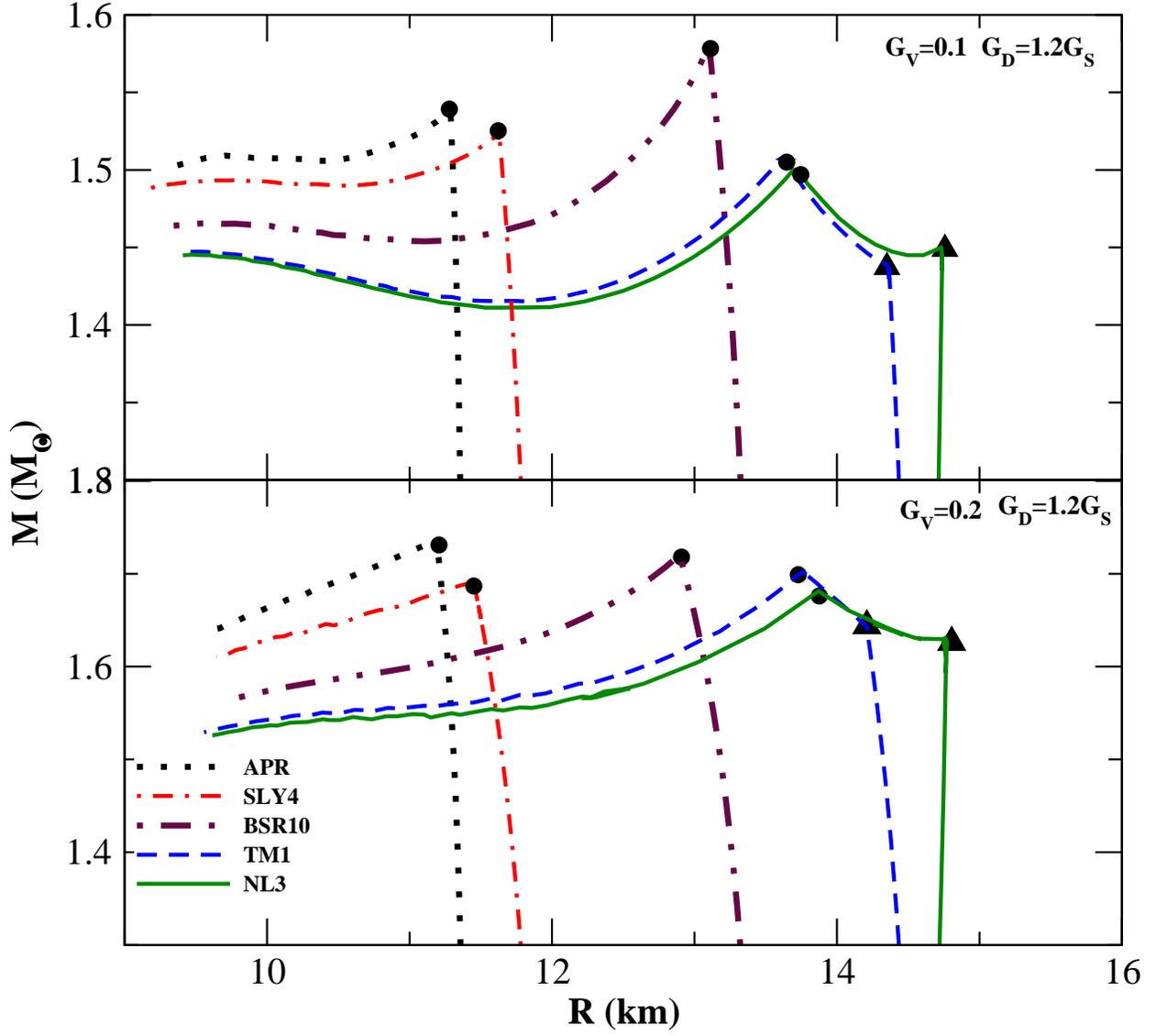}}
\caption{\label{fig:fig5} (Color online) Plots for the mass-radius
relationships for the equilibrium sequences of the  non-rotating compact
stars obtained using various EOSs as shown in Fig. \ref{fig:fig3}. The
solid circles and triangles divide the curves according to the composition
of the hybrid stars as described in Fig. \ref{fig:fig4}.  }
   \end{figure}

The equilibrium sequence of compact stars  for a given  EOS is obtained
by varying the central energy density $\epsilon_c$. For the stable
configuration,
 \begin{equation} \frac{\partial M}{\partial \epsilon_c} \geqslant 0,
\label{eq:sbl} \end{equation}
 where, $M$ is the gravitational mass of the non-rotating
compact star.  The equilibrium sequences for the non-rotating compact
stars resulting from our EOSs are plotted as mass versus radius in
Figs. \ref{fig:fig4} and \ref{fig:fig5}.  The central energy density
increases as we move along these curves from the right hand side.
The portion of the curves left to  the solid circles represent the
equilibrium sequences of hybrid stars with CFL quark matter core.
The curves between the solid circle and triangle represent the sequences
of hybrid star composed of 2SC quark matter core.  It is clear from
the lower panel of Fig. \ref{fig:fig4} and upper panel of Fig. \ref{fig:fig5}
that the stable configurations of hybrid stars with CFL quark matter
core belong to third family of  compact stars. Further, irrespective of the
choice of the EOS of the nuclear matter,  the stable
configurations of the non-rotating hybrid stars exist within the
NJL model only when the EOSs for the CSQM are constructed for $G_D =
1.2G_S$ with $G_V \lesssim 0.1G_S$.  These values of $G_D$ and $G_V$, for
which the stable configurations of the hybrid star exist, are very much
similar to the ones found in Ref.  \cite{Pagliara08}. It appears that the
stability of the hybrid stars with CSQM core depends solely on the
choice of the EOS for the CSQM.  However, the composition of the hybrid
stars depend on the behaviour of the  nuclear matter EOS.  For instance,
in case of the TM1 and the NL3 EOSs of the nuclear matter, the core of the
hybrid stars are composed of CSQM which is either in 2SC phase or in
the CFL phase. In the later case, CFL quark matter core is surrounded by
a layer of 2SC quark matter with the outer layer composed of nuclear
matter . The thickness of the 2SC quark matter at the maximum hybrid star
mass is around $0.5 - 0.7$km and its mass is $\sim 0.1M_\odot$.  On the
other hand, no 2SC quark matter appears in the stable configurations of
the hybrid star constructed using the EOSs for which the nuclear matter
part correspond to the APR and SLY4.

\begin{figure}[ht]
 \centering
\resizebox{6.5in}{!}{\includegraphics[]{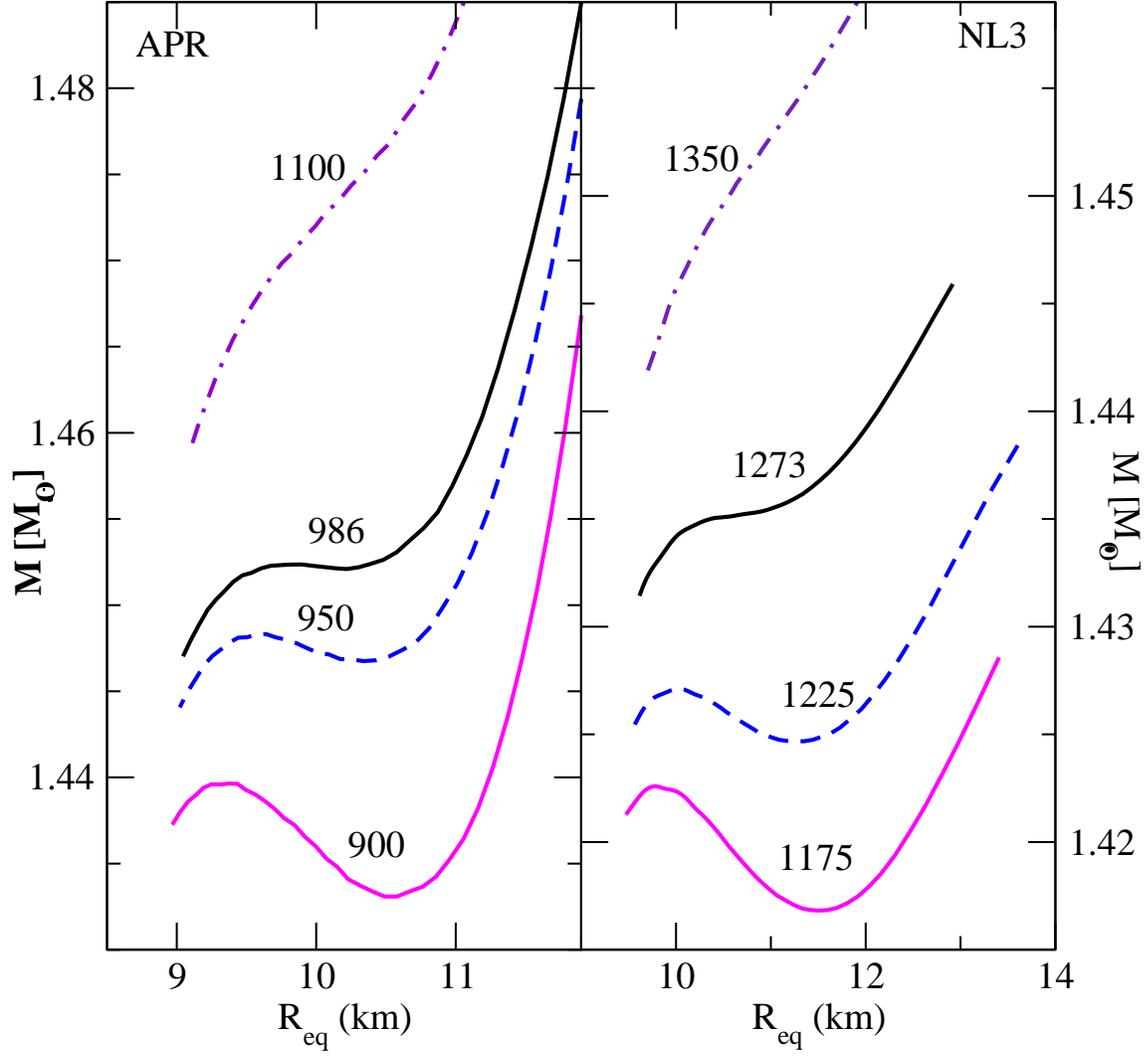}}
\caption{\label{fig:fig6} (Color online)
Plots for the mass verses circumferential equatorial radius $R_{\rm eq}$
at fixed values of the rotational frequency as indicated along each of
the curves (in Hz). The black solid lines represent the results obtained
at the critical frequencies $f_{\rm crit}$.  For $f>f_{\rm crit}$,
the stable configurations of hybrid star do not exist.  }
   \end{figure}

\begin{figure}[ht]
 \centering
\resizebox{6.5in}{!}{\includegraphics[]{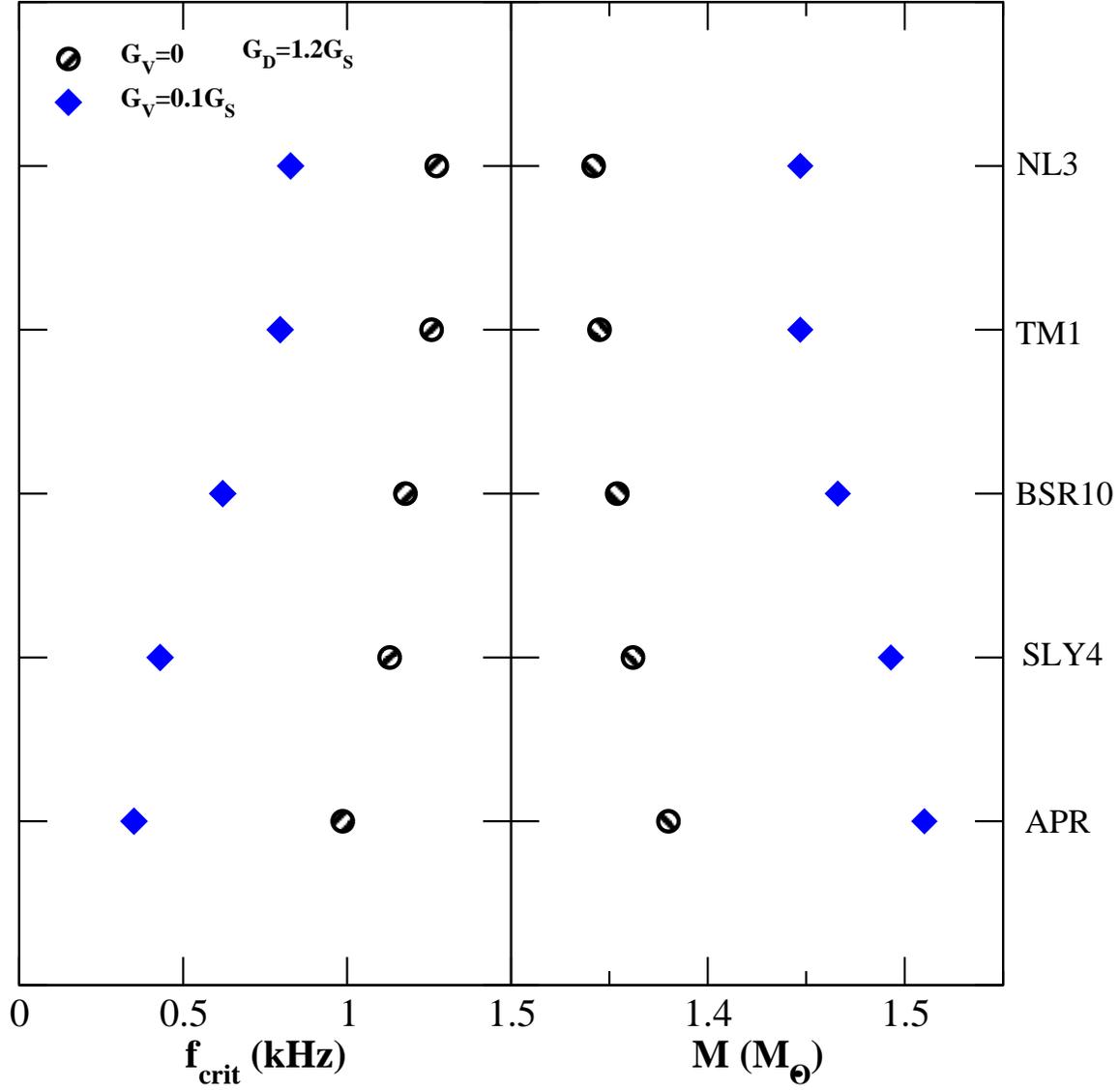}}
\caption{\label{fig:fig7} (Color online)
The values of the critical rotation frequency $f_{\rm crit}$ (left panel)
for the hybrid stars with the CSQM core and their maximum masses
(right panel) in the non rotating limit as obtained using different EOSs for
the nuclear matter and the CSQM.
}
   \end{figure}

The equilibrium sequences for the hybrid stars rotating with fixed
rotation frequency $f$ are constructed. As an  illustration, in Fig.
\ref{fig:fig6}, we plot mass verses circumferential equatorial radius
$R_{\rm eq}$ at fixed values of the rotational frequency obtained  for
two different EOSs.  For the clarity, we mainly focus on the regions of
the $M - R_{eq}$ curves corresponding to the sequences of the hybrid
stars which  are relevant in the present context.  We see that beyond
certain frequency, so-called the critical rotation frequency $f_{\rm
crit}$, the stable configuration for the rotating hybrid star  does not
exist. The solid black lines in Fig. \ref{fig:fig6} represent the result
obtained at the $f = f_{\rm crit}$.  In Fig. \ref{fig:fig7} we plot the
values for the $f_{\rm crit}$ (left panel) calculated for the cases for
which the stable configurations for the non-rotating hybrid star exist.
It is evident that the values of $f_{crit}$  are quite sensitive to the
choice of the EOS for the nuclear matter as well as the CSQM. Depending
on the EOSs, $f_{crit}$ varies in the range of $350 - 1275$ Hz.  We also
plot in Fig. \ref{fig:fig7} (right panel) the maximum mass $M_{\rm max}$
for the non-rotating hybrid stars with CSQM core.  It is interesting
to note from this Fig. that the values of $G_D$ and $G_V$ for a given
nuclear matter EOS can be so adjusted that the resulting hybrid star has
(a) the maximum mass in the non-rotating limit larger than $1.44M_\odot$
which is the most accurately measured value for the maximum mass of a
compact star \cite{Thorsett99} and (b) the maximum allowed rotation
frequency is larger than the current observational limit of $716$
Hz \cite{Hessels06}.

\begin{figure}[ht]
 \centering
\resizebox{6.5in}{!}{\includegraphics[]{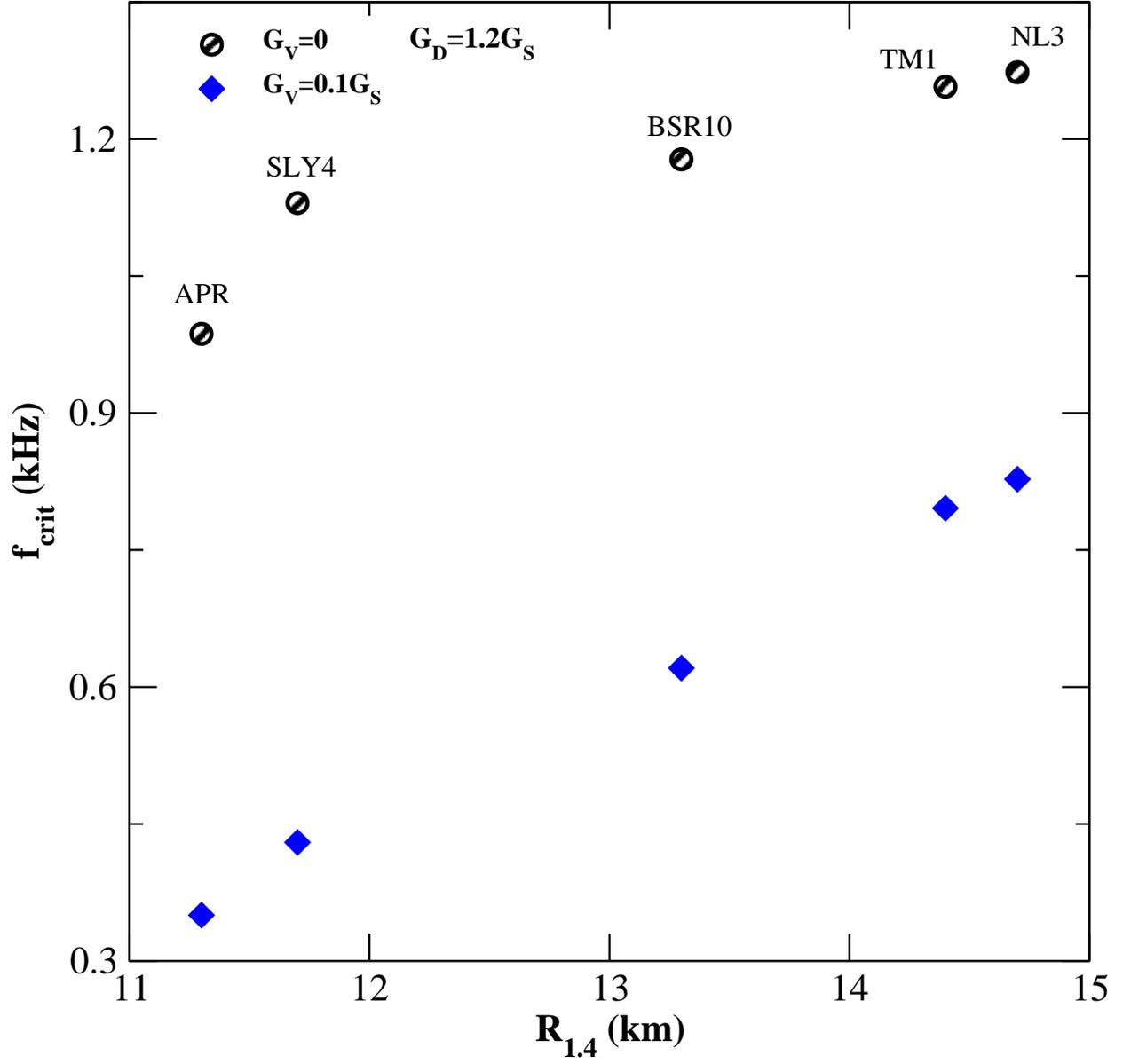}}
\caption{\label{fig:fig8} (Color online)
Correlations between  values of $f_{\rm crit}$ for the hybrid stars 
and the radius $R_{1.4}$ for the neutron star with
canonical mass  as listed in Table \ref{tab:tab1}.
}
   \end{figure}

In Fig. \ref{fig:fig8} we show the correlations between the values of
the $f_{\rm crit}$ for the hybrid stars and the radius $R_{1.4}$ for the
neutron star with the canonical mass. We see that $f_{\rm crit}$ is large
if value of $R_{1.4}$ is also large. Thus, hybrid star constructed for
a given EOS for the CSQM can rotate faster if the EOS for the nuclear
matter is stiffer. The existence of the correlations between the values
of $f_{\rm crit}$ and $R_{1.4}$ may be due to the fact that the  pressure
at which the nuclear to the quark matter transition occurs is closer to
the values of $P_{1.4}$ as can be seen from the lower and upper left
panels of Figs. \ref{fig:fig2} and \ref{fig:fig3}, respectively.

Finally,  we would like to compare the present results
with coresponding ones obtained within the MIT bag model
\cite{Alford05,Agrawal09}. The present results as obtained
within the NJL model are significantly different with those  for the MIT
bag model.  Within  MIT bag model the EOS for the CSQM can be obtained
by adjusting the value of the CFL gap parameter and the bag constant
such that the resulting hybrid stars with CFL quark matter core are
gravitationally stable upto the masses $\sim 2M_\odot$ in the static
limit and the the maximum allowed rotation frequency is much larger than 1 kHz.
However, it can be seen from Fig. \ref{fig:fig7}, stable configurations
of the hybrid stars with CFL quark matter core obtained within the NJL
model are having the maximum values for the mass and the rotational
frequency appreciably lower than those obtained for the MIT bag model.
The differences between the results for the MIT bag model and the NJL
model can be attributed to the fact that the constituent quark masses,
chiral condensates  and the colour superconducting gaps in the later case
are computed self-consistently as a function of baryon density.

\section{Conclusions}
The stability of  non-rotating and rotating hybrid stars, composed of the
colour superconducting quark matter core surrounded by a  nuclear mantle,
is studied by using several EOSs.  The EOSs for the nuclear matter, employed
at lower densities,  are based on the variational and the mean field
approaches.  We use a diverse set  of nuclear matter EOSs such 
that the resulting maximum neutron star mass lie in the range of $2.2
- 2.8M_\odot$ and the radius at the canonical neutron star mass vary
between $11.3 - 14.8$km.  The EOSs at higher densities corresponding
to the colour superconducting quark matter in the 2SC or the CFL phase,
are calculated within the NJL model using different values of coupling
strengths for the scalar diquark and isoscaler vector terms.  The EOS
at intermediate densities are obtained using a Maxwell construction.

We find that the stability of the non-rotating hybrid stars is very
much  sensitive to the EOSs for the colour superconducting quark matter and
almost independent of the choice for the EOS for the nuclear matter.
The stable configurations of the hybrid stars exist only for the large
enough value for the scalar diquark coupling strength.  Though, the
stability of the hybrid stars are not sensitive to the choice of the EOSs
for the nuclear matter,  but, compositions of the hybrid stars are at
variance for these EOSs.  If the EOS for the nuclear matter is stiff,
core of the hybrid star is composed of colour superconducting quark
matter  which is either in the 2SC or the CFL phase. In the later case,
CFL quark matter core is surrounded by a thin layer of the 2SC quark matter and
the outer layer composed of nuclear matter .

The stability of the rotating hybrid star is sensitive to the choice
of the EOS for the nuclear matter as well as that for the colour
superconducting quark matter. In particular, we find that the values
of the critical rotation  frequency vary from about $350$ Hz to $1275$
Hz depending upon the choice of the EOSs for the nuclear matter and the
colour superconducting quark matter.  Our results also indicate that the
EOSs for the colour superconducting quark matter obtained within the NJL
model may be adjusted for the various nuclear matter EOSs in such a way
that it yields (a) the maximum mass in the non-rotating  limit larger
than $1.44M_\odot$ which is the most accurately measured value for the
maximum mass of a compact star and (b) the maximum allowed rotation
frequency is larger than the current observational limit of $716$ Hz.

Finally, we would like to  mention that our present study can be
extended in several ways.  The quark matter in the crystalline color
superconducting phase, expected to appear at the intermediate densities,
should also be considered.   One might also include the contributions from
the hyperons which would soften the hadronic EOS.  The phase transition
from hadron to the quark matter should proceed via mixed phase which
can be constructed using the Gibbs conditions.

\newpage

\newpage
\begin{table}
\caption{\label{tab:tab1}
Values of the maximum mass $M_{\rm max}$  and corresponding central
energy density $\epsilon_{\rm max}$  and radius $R_{\rm max}$ obtained
for different EOSs of the nuclear matter. The radius $R_{1.4}$ for the
neutron star with canonical mass ($1.4M_\odot$) are also given.}
\begin{tabular}{|c|c|c|c|c|c|c|}
\hline
 EOS& Approach  &$\epsilon_{\rm max}$& $M_{\rm max}$& $R_{\rm max}$&
$R_{1.4}$&Ref.\\
&& $(10^{15}\text {g/cm}^3)$& $(M_\odot)$& (km)& (km)&\\
\hline
APR& Variational&2.80 & 2.19& 9.9&11.3& \cite{Akmal98}  \\
SLY4& NRMF&2.84& 2.05& 10.0& 11.7 & \cite{Douchin01} \\ 
BSR10
\footnote{This EOS 
is obtained using one of the several parameter sets of the
extended RMF model given in our earlier work \cite{Dhiman07}.  Each of
these parameterizations corresponds to different values of the strength
$\zeta$ for the  $\omega$ -meson self-coupling term and neutron-skin
thickness $\Delta r$ in $^{208}$Pb nucleus.  The remaining parameters
of the models were calibrated to yield reasonable fit to the bulk
nuclear observables and nuclear matter incompressibility coefficient.
In the present work we use the  parameter set with $\zeta = 0.03$ and
$\Delta r = 0.2$fm which will be referred henceforth as BSR10.}
& RMF& 2.14& 1.97& 11.6& 13.3 &  \cite{Dhiman07}\\
TM1& RMF&1.87 &2.19 &12.4 & 14.4& \cite{Sumiyoshi95} \\
NL3& RMF&1.55 &2.79 &13.3 &14.7 & \cite{Lalazissis97} \\
\hline
\end{tabular}

\end{table}

\newpage
\newpage

\newpage 

\newpage

\newpage

\end{document}